\documentclass{article}
\usepackage{spconf,amsmath,epsfig}
\usepackage{color}
\usepackage{xcolor}
\usepackage{amsthm,amssymb}
\usepackage{comment}

\newcommand{\done}{\rlap{$\square$}\raisebox{.15ex}{\hspace{0.1em}$\checkmark$}%{\raisebox{2pt}{\large\hspace{1pt}\cmark}}%
\hspace{-1pt}}

\title{Assimilation of SWOT Altimetry and Sentinel-1 Flood Extent Observations for Flood Reanalysis - A Proof-of-Concept}
\name{T. H. Nguyen$^{1,2}$, S. Ricci$^1$, A. Piacentini$^1$, C. Emery$^3$,  R. Rodriquez Suquet$^4$, S. Peña Luque$^4$\thanks{Thanks to SCO and TOSCA agencies for funding.}}
\address{\small
$^1$ CECI-UMR 5318 CERFACS-CNRS, Climate Modeling and Global Change, 31057 Toulouse Cedex 1, France\\
\small $^2$ Luxembourg Institute of Science and Technology, 4362 Esch-sur-Alzette, Luxembourg\\
\small $^3$ CS-Group, 31401 Toulouse Cedex 9, France\\
\small $^4$ Centre National d'Études Spatiales (CNES), Earth Observation Laboratory, 31401 Toulouse Cedex 9, France 
}

\begin{document}
%\ninept
%
\maketitle
\begin{abstract}
%Floods are one of the most common and devastating natural disasters worldwide. 
In spite of astonishing advances and developments in remote sensing technologies, meeting the spatio-temporal requirements for flood hydrodynamic modeling remains a great challenge for Earth Observation. The assimilation of multi-source remote sensing data in 2D hydrodynamic models participates to overcome such a challenge. The recently launched Surface Water and Ocean Topography (SWOT) wide-swath altimetry satellite provides a global coverage of water surface elevation at a high resolution. SWOT provides complementary observation to radar and optical images, increasing the opportunity to observe and monitor flood events. This research work focuses on the assimilation of 2D flood extent maps derived from Sentinel-1 C-SAR imagery data, and water surface elevation from SWOT as well as in-situ water level measurements. An Ensemble Kalman Filter (EnKF) with a joint state-parameter analysis is implemented on top of a 2D hydrodynamic TELEMAC-2D model to account for errors in roughness, input forcing and water depth in  floodplain subdomains. The proposed strategy is carried out in an Observing System Simulation Experiment based on the 2021 flood event over the Garonne Marmandaise catchment. This work makes the most of the large volume of heterogeneous data from space for flood prediction in hindcast mode paves the way for nowcasting.
%“Flood forecasting” pertains to the estimation of future floods, while “flood prediction” is typically used to describe the characterization of past flood events (hindcasting) or provid near-real-time information (nowcast- ing). I
\end{abstract}
\section{Introduction}
\label{sec:intro}

The inclusion of Earth Observations (EO) from space into flood risk prediction strategies presents a great opportunity to improve the ability to monitor, anticipate flooding, mitigate its impacts, as well as protecting assets worldwide. The properties of inland water bodies are monitored by altimetry missions that provide along-track water surface elevation (WSE) from nadir (e.g., TOPEX/Poseidon, Jason, SARAL/AltiKa, Sentinel-3, Sentinel-6) or large-swath altimeters (SWOT), as well as from radar/optical satellites (Sentinel-1/Sentinel-2) that provide high-resolution water extent maps.
Nevertheless, remote-sensing (RS) data can provide spatially dense information but with a sparse revisit frequency. On the other hand, in-situ gauge measurements offer a high temporal sampling but they are limited to low spatial coverage. Neither suffice to fully capture both spatial and temporal dynamics of flooding. 
% The SWOT wide-swath altimetry satellite provides a global coverage of WSE at a high resolution presented as a rasters and pixel clouds from which river products are computed along the predefined river centerlines.
% Data assimilation allows to combine physics-based equations from hydrodynamic models with ground measurements and multi-mission satellite data that have different characteristics in space and time.

Data assimilation (DA) has emerged as an efficient tool to deal with uncertainties in hydrology and hydrodynamics, such as in roughness, inflow discharge, channel and floodplain geometry and/or hydraulic state. It combines physics-based equations that govern the hydrodynamic models with ground measurements and/or multi-mission satellite data that offer different characteristics in space and time. Therefore, it allows to correct initial, boundary conditions and model parameters to issue improved nowcasts and forecasts. Leveraging multi-source flood observations allows densifying the observing network, both spatially and temporally, as well as diversifying their characteristics. Previous works \cite{NguyenTGRS2022,nguyenagu2022, nguyen2023gaussian} have shown that this allows for a better performance of the EnKF---that relies on the stochastic computation of forecast error covariance matrix among a limited number of perturbed simulations---to represent the flow dynamics in the riverbed and floodplain. Indeed, flood extent maps derived from Sentinel-1 (S1) C-band SAR imagery data were assimilated jointly with in-situ water depth measurements from observing stations of the VigiCrue network in France, over the Garonne Marmandaise catchment; this improves the representation of the major flood events in 2019 and 2021 \cite{nguyenagu2022, nguyen2023gaussian}. 
Launched in December 2022, the SWOT wide-swath altimetry satellite, with the Ka-band Radar Interferometer onboard, shall provide a global coverage of WSE at a high resolution presented as rasters and pixel clouds from which river products are delivered along the predefined river centerlines.

The present work investigates the addition, to this workflow, of river products derived from SWOT satellite at nodes located over the centerline of the river. Given the recent launch of SWOT and the limited availability of real SWOT river product, the experiment is carried out in an Observing System Simulation Experiment (OSSE) framework using synthetical S1, SWOT and in-situ observations generated from a reference simulation.

% \sophie{The paper is organized as follows: Sect.~\ref{sec:method} briefly describes the methodology for synthetical observations generation, numerical modeling and data assimilation. The experimental settings and results are given in Sect.~\ref{sec:result}, followed by brief conclusions and perspectives.}

\section{Method}
\label{sec:method}

\begin{figure}[h]
\centering
\includegraphics[trim=0 0.38cm 0 0, clip, width=\linewidth]{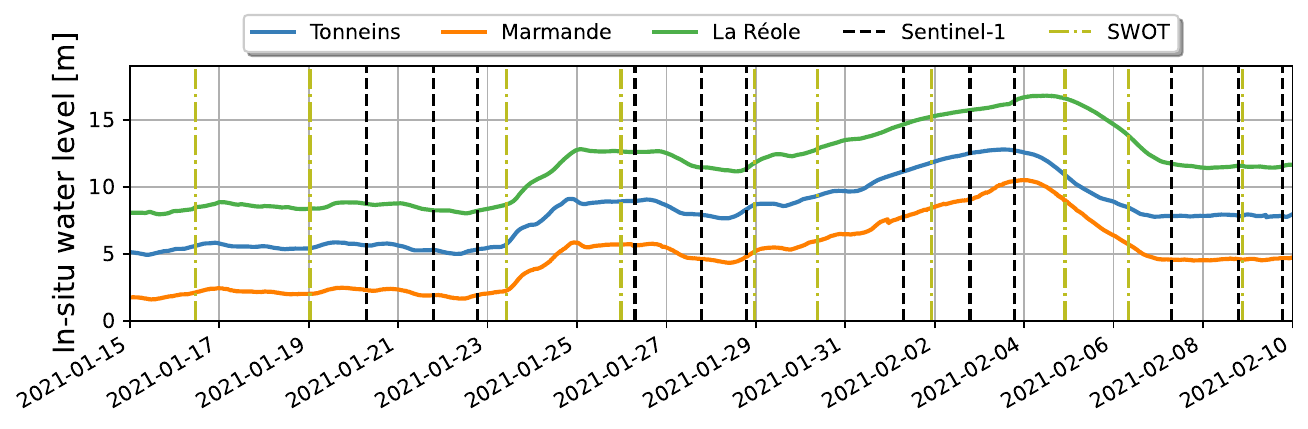}
\caption{Observations for OSSE based on the 2021 event.}
% Synthetical in-situ WSE time-series at Tonneins, Marmande and La Réole are plotted in blue, orange and green. The overpass times for S1 and SWOT are indicated as vertical black dashed lines and yellow dash-dotted lines.}
\label{fig:obs}
\end{figure}

The study is carried out with the TELEMAC-2D\footnote{\texttt{https://opentelemac.org/}} (T2D) hydrodynamic solver, over a 50-km reach of the Garonne River between Tonneins and La Réole, for a major flood event in 2021. The upstream forcing is provided by the observed stage height at Tonneins translated into discharge using a local rating curve.
The OSSE observing setting was based on a real scenario for in-situ and S1 data, whereas the SWOT pass plan was enhanced, compared to the 21-day repeat cycle of the nominal science orbit, by tripling the number of passes during the flood period. This allows a temporal sampling of 1-2 days for SWOT over the studied catchment, which is more ideal for flood studies.
Fig.~\ref{fig:obs} depicts the synthetical in-situ water-level for the 2021 flood event at VigiCrue observing stations, namely Tonneins (blue), Marmande (orange) and La Réole (green). The overpass times of S1 and SWOT are indicated as vertical black dashed lines and yellow dash-dotted lines, respectively.

\begin{figure}[t]
\centering
\includegraphics[width=\linewidth]{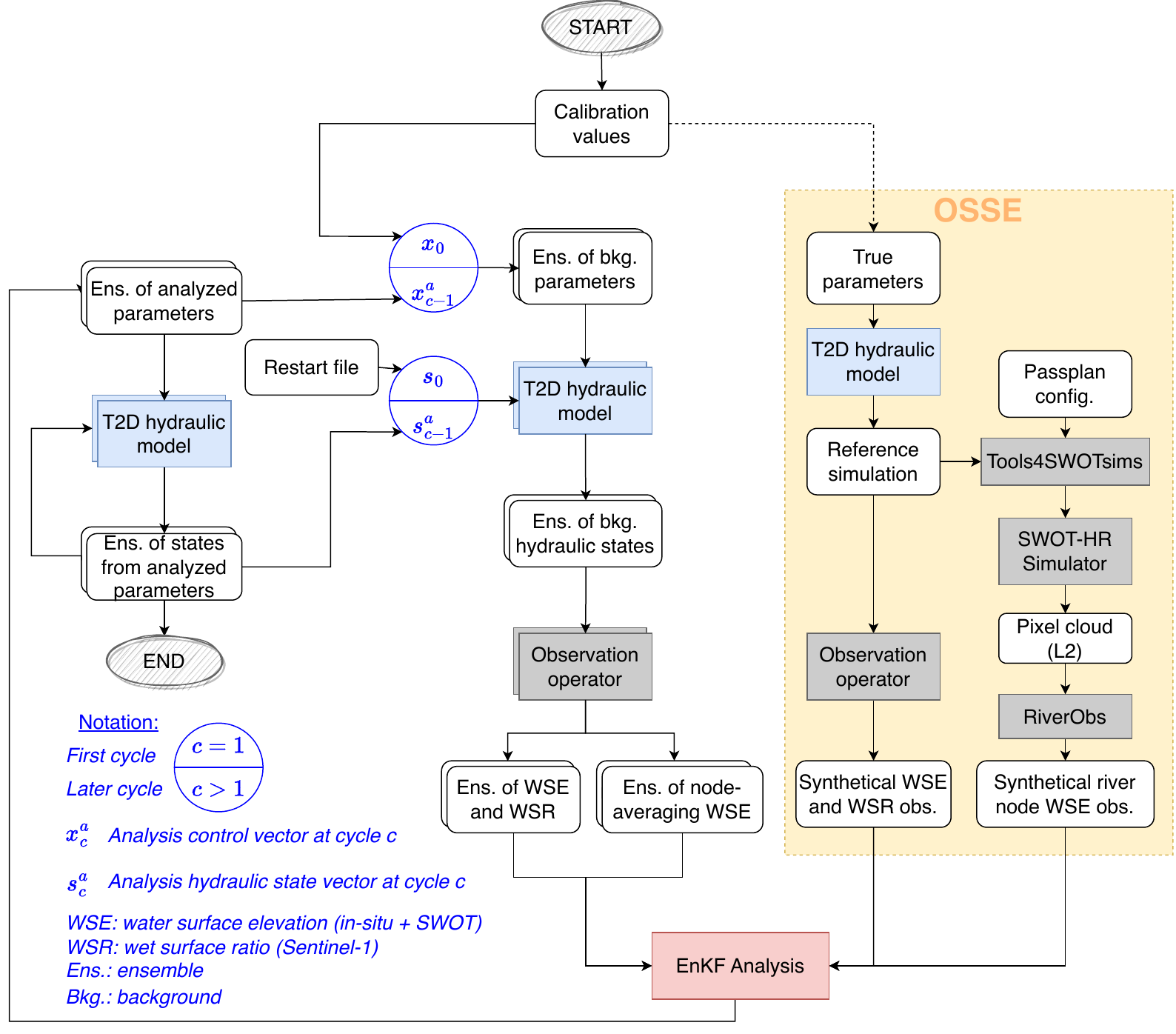}
\caption{Workflow of EnKF DA strategy in OSSE.}
\label{fig:workflow}
\end{figure}

\begin{figure}[!t]
\centering
\includegraphics[trim=0 0cm 0 0,clip,width=\linewidth]{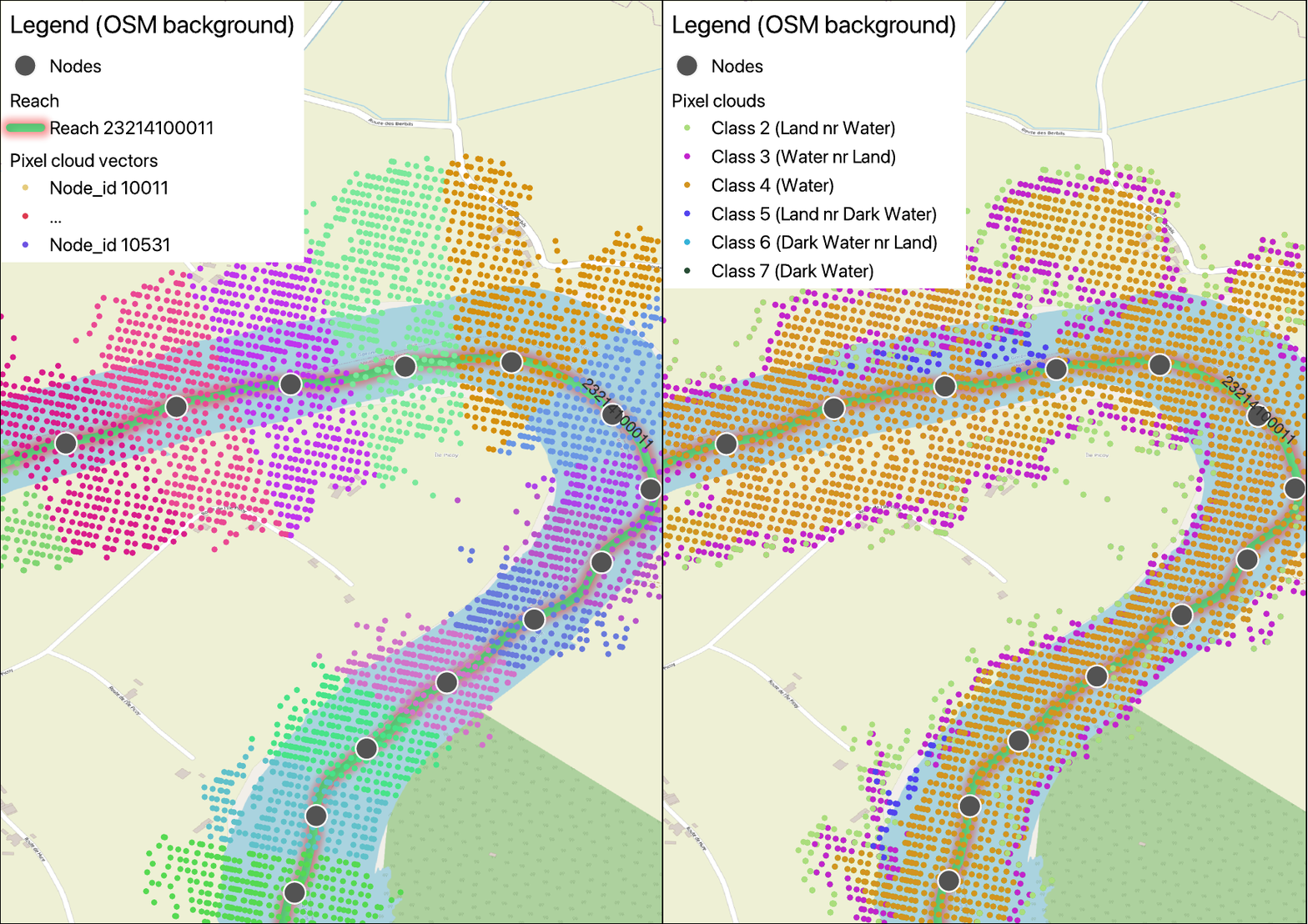}
\caption{Generation of node-level WSE SWOT river products on 2021-01-16. Color-coded aggregated pixels for each node (left panel) and pixel class (right panel). }
\label{fig:SWOT_nodes}
\end{figure}

The data assimilation workflow in OSSE is presented in Fig.~\ref{fig:workflow}.
An EnKF with a joint state-parameter analysis is implemented on top of T2D \cite{nguyenagu2022}. The EnKF control vector is composed of spatially-distributed friction coefficients $K_s$, a corrective parameter of the inflow discharge $Q$, typically necessary when dealing with highly uncertain inflow \cite{nguyen2023reducing}. In addition, it is augmented with a uniform correction of hydraulic states within the floodplain subdomains $\delta H$. The ensemble of T2D simulations for the EnKF algorithm are generated from a priori values for the friction and upstream forcing perturbed with errors with gaussian statistical properties~\cite{NguyenTGRS2022,nguyenagu2022}. The settings of the OSSE's reference simulation are based on a previous assimilation of real in-situ data. This reference simulation are used to generate synthetic observations, thanks to the dedicated observation operators. 
Such operators are used to compute the model equivalent of the observations for each member of the ensemble.
%to replicate the in-situ water level, Sentinel-1-derived flood extent and SWOT WSE observations. 
True WSE values are extracted from the reference simulation at all observation times and locations to generate synthetical in-situ data at stream-gauge stations. On the other hand, a threshold of 5 cm is applied on reference water depth maps to generate synthetical flood extents at S1 overpass times, which are used to compute wet surface ratios (WSR) in selected subdomains of the  floodplain \cite{nguyenagu2022}. Such non-Gaussian ratio is further treated with a Gaussian anamorphosis \cite{nguyen2023gaussian,nguyen2023dealing}.

The synthetical SWOT river product are generated from the reference water depth maps at SWOT overpass times, first processed with the Tools4SWOTsims toolbox \cite{emery2022tools4swotsims}. This toolbox consists of a set of Python scripts to map 1D/2D hydrodynamic model outputs into 2D WSE rasters that are compatible with SWOT-HR simulator\footnote{\texttt{https://github.com/CNES/swot-hydrology-toolbox}}, dedicated to hydrology science \cite{frasson2017}. The outputs of the SWOT-HR simulator are SWOT observations in pixel cloud format. Such pixel cloud data is then processed by the RiverObs\footnote{\texttt{https://github.com/SWOTAlgorithms/RiverObs}}
% This package translates two-dimensional imagery to one-dimensional measurements of WSE, width, and slope at nodes located approximately every 200m along river longitudinal profiles defined by the SWOT River Database (SWORD)\footnote{http://gaia.geosci.unc.edu/SWORD/} ~\cite{allen2018} from GRWL dataset and hydraulic reaches \cite{altenau2021}.} 
which aggregates the WSEs from a selection of pixels to issue WSEs at nodes every 200~m along river centerline and at river reach every 10~km (which are defined by the SWOT River Database  \cite{allen2018}).
The selection of pixels is based on a class assignment related to the nature of the water pixels (i.e. open water, water near land, dark water, etc), and weighted by their respective uncertainties to issue node-level WSE with an error below 10~cm as prescribed in the SWOT requirements.
The details for SWOT WSE, width, and slope computation are given in  SWOT River Single Pass Product Description Document \cite{swotproduct2020} 
% (JPL Internal Document, 2020) 
and example data products\footnote{\texttt{https://podaac.jpl.nasa.gov/swot?tab=datasets}}.

Fig.~\ref{fig:SWOT_nodes} depicts a typical SWOT node-level river product over a meander of a reach of the Garonne River observed on 2021-01-16, each node is an aggregation of pixels. The pixels are color-coded either by their associated node on the left panel, or by their class on the right panel. For each member of the ensemble, the model equivalent of the SWOT node-level WSE is computed using the SWOT-dedicated observation operator. 
It is the average of the WSEs of certain associated pixels selected according to their quality flag; each pixel WSE is interpolated from the T2D mesh of simulated water depth.
% It aggregates the WSEs simulated over the nodes from the T2D unstructured mesh that are associated (2D-interpolation) with the selected pixels.

\section{Experimental Results}
\label{sec:result}

Four EnKF DA experiments are carried out considering different combinations of in-situ, S1 and SWOT synthetical observations, as presented in Table~\ref{tab:settings}. In the following, the deterministic simulation using the control vector true values is represented as \textit{Truth}, and the simulation without DA using the calibrated parameters is called  \textit{Open Loop} (OL). IDA assimilates only in-situ water-level measurements, IGDA assimilates in-situ WSE and S1-derived WSR observations, FDA assimilates all available flood observations whereas RSDA assimilates only remote-sensing data. 

Quantitative performance assessments are carried out in the observational spaces by comparing the DA analysis WSE time-series at observing stations with the true water-level observations, and by the means of the resulting contingency maps and the overall Critical Success Index (CSI) metric computed for the DA analysis flood extent maps, with respect to the \textit{truth} flood extent maps.

\begin{table}[h]
\caption{Summary of experimental settings.}
\label{tab:settings}
\begin{tabular}{p{0.1\linewidth}cccc}
    \hline
    & \multicolumn{3}{c}{Assimilated observations} &   \\ \cline{2-4}
    Exp.  & In-situ  & Sentinel-1  & SWOT  &  \\ 
    name & WSE & WSR & WSE & Control vector\\ \hline
%    \textcolor{black}{TRUTH} & $\square$ & $\square$ & $\square$ & - \\
    \textcolor{blue}{OL} & $\square$ & $\square$ & $\square$ & - \\
    \textcolor{green}{IDA} & $\done$ & $\square$ & $\square$ & $K_s, Q$ \\ %\hline
    \textcolor{red}{IGDA} & $\done$ & $\done$ & $\square$ & $K_s, Q, \delta H$ \\ 
    \textcolor{cyan}{RSDA} & $\square$ & $\done$ & $\done$ & $K_s, Q, \delta H$ \\ 
    \textcolor{violet}{FDA} & $\done$ & $\done$ & $\done$ & $K_s, Q, \delta H$ \\
\hline
\end{tabular}
\end{table}

\begin{figure}[!h]
\centering
\includegraphics[trim=0 0.42cm 0 0, clip,width=\linewidth]{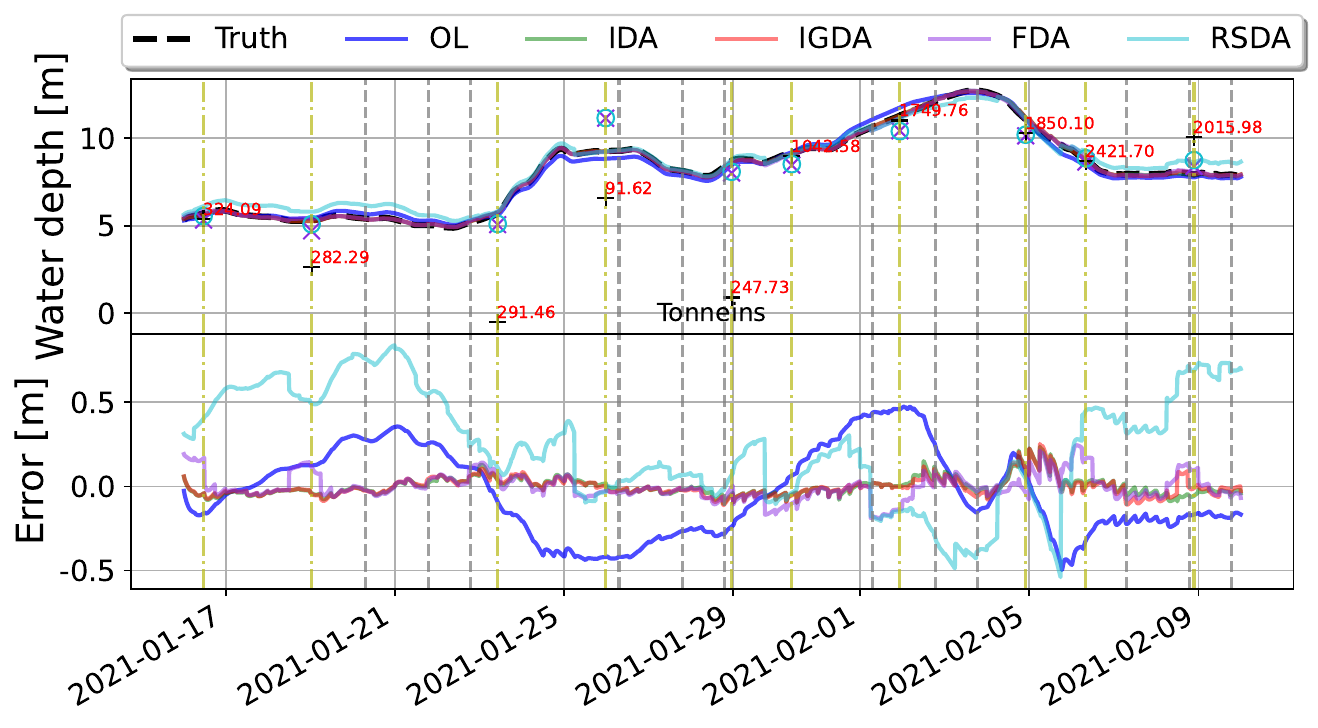}
\caption{WSEs at Tonneins for all experiments (top panel), and errors computed with respect to the \textit{Truth} WSE (bottom panel).}
\label{fig:H_Tonneins}
\end{figure}

\begin{comment}
\begin{figure}[h]
\centering
\includegraphics[width=\linewidth]{fig/H_and_Err_H_Marmande_Vigicrue.pdf}
\caption{WL at Marmande.}
\label{fig:H_Tonneins}
\end{figure}

\begin{figure}[h]
\centering
\includegraphics[width=\linewidth]{fig/H_and_Err_H_La_Reole.pdf}
\caption{WL at La Réole.}
\label{fig:H_Tonneins}
\end{figure}
\end{comment}

Fig.~\ref{fig:H_Tonneins} depicts the WSE simulated at Tonneins from the five experiments on the top panel, and the anomalies to the truth on the bottom panel. The SWOT-derived WSE at the closest nodes to Tonneins are represented by black crosses. The RMSEs of the WSE at all three observing stations computed over the entire flood event are provided in Table~\ref{tab:rmse}. The assimilation of in-situ data with high temporal sampling suffices to represent the WSE (in the riverbed) at the gauge stations. It reduces the RMSE from 26~cm in the OL to 5-7~cm in IDA, IGDA and FDA. 

\begin{figure*}[t]
\centering
\includegraphics[width=0.8\linewidth]{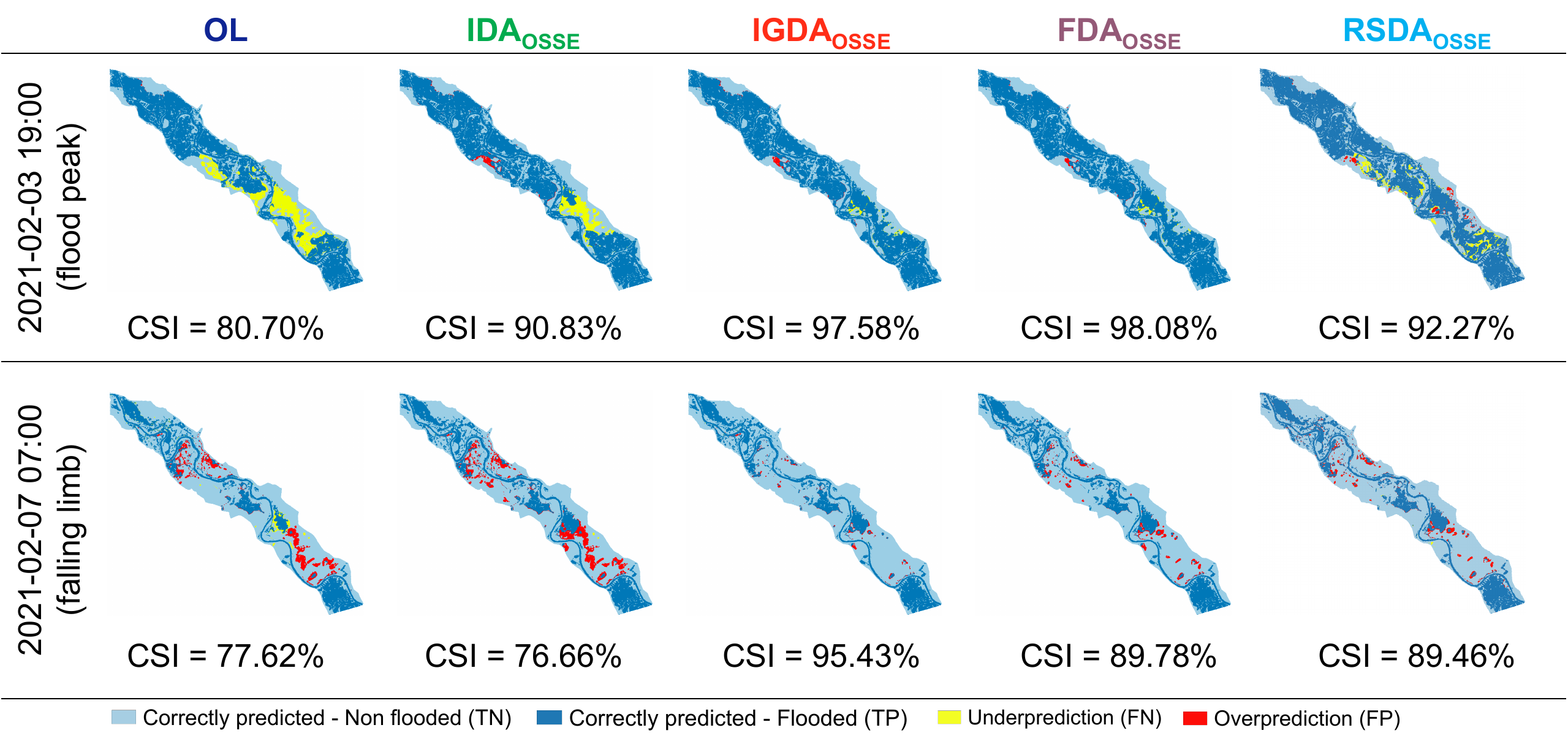}
\caption{Contingency maps for all experiments with respect to \textit{truth} flood extents at flood peak and recess.}
\label{fig:conti}
\end{figure*}

The improvement brought by FDA demonstrates the feasibility of multi-sources DA. It should be noted that for RSDA, in-situ observations are independent observations. The discrepancies between in-situ and SWOT WSE are most likely due to the space aggregation of SWOT pixels, thus the assimilation of SWOT does not guarantee an improvement with respect to the WSEs at gauge stations. In RSDA, the correction at Tonneins is implicitly influenced by misfits to SWOT at various locations, including those precede Tonneins. Here, the assimilation of SWOT degrades the results with respect to in-situ observations at Tonneins. These results from RSDA should be further investigated, with a focus on the consistency between the SWOT node-level WSE and the \textit{truth} WSE.
%The poor temporal sampling of SWOT data does not allow to efficiently reduce the RMSE with respect to in-situ observations, it even degrades the simulation between the SWOT overpasses. }
 The contingency maps and CSI computed with respect to the \textit{truth} flood extents are shown in Fig.~\ref{fig:conti} for all experiments. OL tends to heavily underestimate the flooding at the flood peak and overestimate the flooding during recess. The assimilation of synthetical S1 WSR and/or SWOT node-level WSE (in IGDA, RSDA and FDA) improves the representation of the floodplain dynamics and thus have better CSI score with respect to OL and IDA.

% \clearpage
\begin{table}[h]
\caption{Root-mean-square errors of simulated WSE at observing stations.}
\label{tab:rmse}
\centering
\begin{tabular}{cccc}
\hline
RMSE [m] & Tonneins & Marmande & La Réole\\ \hline
\textcolor{blue}{OL} & 0.260 & 0.398 & 0.578 \\ %\hline
\textcolor{green}{IDA} & 0.052 & 0.042 & 0.053 \\ %\hline
% IWDA & 0.069 & 0.077 & 0.081 \\ %\hline
% IHDA & 0.065 & 0.073 & \underline{0.079} \\ %\hline
\textcolor{red}{IGDA} & 0.054 & 0.044 & 0.049  \\ 
\textcolor{cyan}{RSDA} & 0.410 & 0.435 & 0.413 \\ 
\textcolor{violet}{FDA} & 0.075 & 0.056 & 0.053 \\ 
\hline
\end{tabular}
\end{table}

\section{Conclusions and Perspectives}\vspace{-0.35cm}
This paper demonstrates the feasibility of multi-sources DA and presents the merits of assimilating different types of flood observations, in the framework of an EnKF built on top of a T2D hydrodynamic model. The FDA experiment that assimilates all observations presents moderate performance, suggesting potential inconsistency between the in-situ and SWOT-nodes observations.
It also shows that the assimilation of remote-sensing only observations (in RSDA) fails to improve the OL results in the riverbed, in exchange for improving the floodplain dynamics. This trade-off will be investigated in further studies with a spatial analysis along the river SWOT nodes. 
Lastly, the good performance of the joint assimilation (in IGDA) of in-situ WSE data and S1-derived flood extent maps demonstrates the benefits of these two data sources, which are the ones with the lowest temporal sampling in real-world applications.
This work heralds toward a reliable methodology for flood prediction for poorly-gauged catchments, making the most of innovative EO data such as real SWOT data.

%Quantitative 1D and 2D metrics are computed to assess the merits of the EnKF strategy on the estimation of water level time-series at gauge stations in the river bed and the representation of the flood extents in the floodplain.
%RS and in-situ data complement well as they present opposite characteristics in terms of frequency and spatial coverage, especially as RS provide data in the floodplain. Quantitative assessments based on 1D/2D metrics show promising results. 
%%%%%%%%%%%%%%


\bibliographystyle{ieeetr}
% {\footnotesize
\bibliography{refs}
% }
\end{document}